\documentclass[fleqn,usenatbib]{mnras}

\usepackage{newtxtext,newtxmath}

\usepackage[T1]{fontenc}

\DeclareRobustCommand{\VAN}[3]{#2}
\let\VANthebibliography\thebibliography
\def\thebibliography{\DeclareRobustCommand{\VAN}[3]{##3}\VANthebibliography}

\usepackage{graphicx}	
\usepackage{amsmath}	
\usepackage[UKenglish]{babel}
\title[Rotational (de)excitation of HMgC$_3$N by He]{Collisional excitation of HMgC$_3$N by He: rate coefficients and astrophysical implications}

\author[L. M. Tian et  al.]{
Liman Tian$^{1}$,
Ziyu Xiang$^{1}$,
Wenli Zou$^{2}$
and Xiaolin Chen$^{1}$\thanks{E-mail: chenxl13@zafu.edu.cn}
\\
$^{1}$College of Optical, Mechanical and Electrical Engineering (College of Intelligent Equipment Engineering), Zhejiang A\&F University, Hangzhou ,311300, PR China\\
$^{2}$Institute of Modern Physics, Northwest University, Xi'an, 710127, PR China\\
}

\date{Accepted XXX. Received YYY; in original form ZZZ}

\pubyear{2026}

\begin{document}
\label{firstpage}
\pagerange{\pageref{firstpage}--\pageref{lastpage}}
\maketitle
\thispagestyle{empty}
\begin{abstract}
Reliable collisional data are essential for interpreting the rotational emission of HMgC$_3$N under the non-equilibrium conditions of molecular clouds. We investigate the rotational excitation and de-excitation of HMgC$_3$N induced by collisions with He, combining \textit{ab initio} calculations with quantum scattering theory. A two-dimensional potential energy surface (PES) for HMgC$_3$N--He  was constructed using the CCSD(T)-F12A method in conjunction with the aug-cc-pVTZ basis set. Quantum close-coupling (CC) calculation were then performed to obtain state-to-state integral cross sections and temperature-dependent rate coefficients for rotational transitions with $\Delta j=\pm1$, $\pm2$, and $\pm3$, covering rotational levels up to $j=20$ and temperatures from $3$ to $50$~K. The rate coefficients exhibit a clear propensity for $\Delta j=\pm2$ transitions. The resulting collisional data were incorporated into radiative transfer modelling to constrain the physical conditions and column density of HMgC$_3$N from its observed spectral line emission.  A $\chi^2$ minimization yields a best-fitting H$_2$ volume density of $n(\mathrm{H}_2)=1.233\times10^5~\mathrm{cm}^{-3}$ and  column density of $N(\mathrm{HMgC}_3\mathrm{N})=1.831\times10^{12}~\mathrm{cm}^{-2}$. The derived column density is in approximate agreement with the observational estimate of $(3.0\pm0.6)\times10^{12}~\mathrm{cm}^{-2}$. These results provide new collisional data for non-LTE studies of HMgC$_3$N and support a more realistic treatment of its rotational excitation and radiative transfer in molecular clouds.
\end{abstract}




\section{Introduction}
Magnesium (Mg) is an abundant refractory element in the interstellar medium (ISM), and its gas-phase chemistry is relevant to dust formation and the chemical evolution of circumstellar environments \citep{Wang2024}. Although magnesium is strongly depleted from the gas phase in dense interstellar clouds, gas-phase magnesium-bearing molecules have been detected in the circumstellar envelopes of carbon-rich stars. IRC +10216, one of the most extensively studied carbon-rich asymptotic giant branch (AGB) stars, has been a key target in searches for magnesium-bearing molecules. Following the identification of MgNC in this source \citep{Guelin1993,Kawaguchi1993}, several other magnesium-bearing molecules have been detected, including MgCN \citep{Ziurys1995}, HMgNC \citep{Cabezas2013}, MgCCH \citep{Agundez2014}, MgC$_4$H and MgC$_3$N \citep{Cernicharo2019},  and MgC$_5$N \citep{Pardo2021}. These detections have expanded the inventory of circumstellar molecules and motivated chemical models in which radiative association of Mg$^{+}$ with unsaturated carbon-chain species, followed by dissociative recombination, can produce neutral magnesium-bearing molecules.

The subsequent detection of HMgC$_3$N, the hydrogenated analogue of MgC$_3$N, extended the family of known magnesium-bearing carbon-chain molecules. In 2023, \cite{Cabezas2023} reported the identification of HMgC$_3$N in a 31.0--50.3~GHz spectral-line survey of IRC+10216 conducted with the Yebes 40-m radio telescope, based on six rotational transitions---some with minor blending---and derived a rotational temperature of approximately 17~K. In the absence of laboratory spectroscopic data, this identification was guided by rotational constants obtained from high-level quantum chemical calculations. The reported column density of HMgC$_3$N \citep{Cabezas2023} is about 60\% that of MgC$_3$N \citep{Cernicharo2019}, while its column density is higher than those reported for MgCN \citep{Ziurys1995} and HMgNC \citep{Cabezas2013}. These relative abundances provide useful observational constraints for chemical models of magnesium-bearing species.

However, the interpretation of the observed rotational emission remains limited by the lack of collisional excitation data for HMgC$_3$N. Non-local thermodynamic equilibrium (non-LTE) modelling of rotational spectra requires state-to-state collisional rate coefficients; however, such data for HMgC$_3$N in collisions with either H$_2$, the dominant molecular collision partner in many interstellar environments, or He are not available. For the related species MgC$_3$N, \cite{Hachani2024} calculated He-collision rate coefficients using a two-dimensional PES, covering rotational levels up to $N=40$ and temperatures of 5--100~K. In this study, we perform a quantum scattering analysis of the HMgC$_3$N--He collisional system. We construct a two-dimensional PES and use it to calculate state-to-state integral cross sections and thermally averaged rate coefficients for rotational transitions with $\Delta j=\pm1$, $\pm2$, and $\pm3$, covering rotational levels up to $j=20$ and temperatures of 3--50~K. The resulting rate coefficients are then used in radiative transfer modelling to reproduce the observed HMgC$_3$N lines toward IRC+10216 and to constrain the excitation conditions and molecular column density. Thesis paper is organised as follows. Section \ref{sec:PES} describes the construction of  HMgC$_3$N--He PES, and its radial expansion coefficients. Section \ref{Scat cal} presents the rotationally inelastic (de-) cross sections and thermally averaged rate coefficients. Section \ref{sec:astro} discusses the astrophysical implications, and Section \ref{sec:conclusions}  summarises the main findings.

\section{The \textit{Ab Initio} Computed Interaction Potential}
\label{sec:PES}
\subsection{Potential energy surface}
A high-precision interaction PES for HMgC$_3$N--He was constructed within the rigid-rotor approximation, in which the intramolecular bond lengths of HMgC$_3$N are fixed and only its rotational degrees of freedom are considered. For HMgC$_3$N, the calculated vibrational frequency of $\omega_e = 86~\mathrm{cm}^{-1}$ indicates that vibrational excitation is negligible under the low-temperature conditions considered here (e.g., $\exp(-86/k_{\mathrm{B}}T) \approx 0.2\%$ at $20$ K). As an example, previous work on AlO$^{+}$ ($\omega_e \approx 1000$~cm$^{-1}$) \citep{Denis2018} has shown that its vibrational cross sections remain at least two orders of magnitude lower than pure rotational cross sections over the entire collision energy range, even up to 5000~cm$^{-1}$, confirming the dominance of rotational scattering. This well justifies the use of the rigid-rotor approximation for HMgC$_3$N--He. This approach has also been adopted for related systems such as HNSi/HSiN--He \citep{Tian2026-1} and CaO/CaS--He \citep{Tian2026} .

\begin{figure}
	\centering
	\includegraphics[width=\hsize]{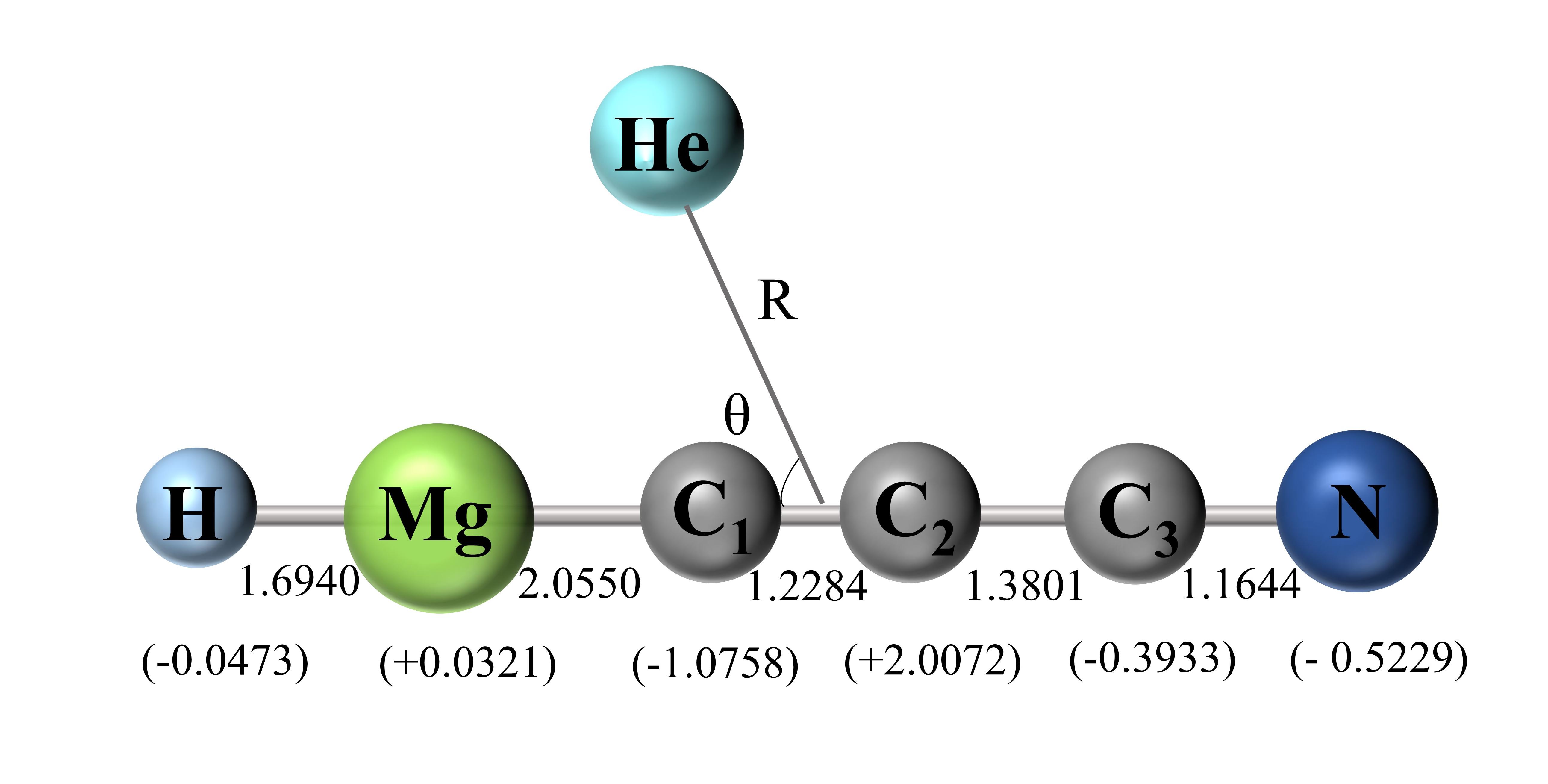}
	\caption{Jacobi coordinate system for the HMgC$_3$N--He.}
	\label{fig.1}
\end{figure}
The PES was constructed in Jacobi coordinates \((R,\theta)\), where \(R\) is the distance between the He atom and the centre of mass of HMgC$_3$N, and \(\theta\) is the angle between the intermolecular vector and the molecular axis. The coordinate system is illustrated in Fig.~\ref{fig.1}. The configuration at \(\theta=0^\circ\) corresponds to the collinear He---HMgC$_3$N arrangement. The equilibrium geometry of HMgC$_3$N was optimised at the CCSD(T)-F12A/aug-cc-pVTZ method. The resulting bond lengths, which were held fixed in the rigid-rotor calculations, are $r_{\text{HMg}} = 1.694$ \AA, $r_{\text{MgC$_1$}} = 2.055$ \AA, $ r_{\text{C$_1$C$_2$}} = 1.2284$ \AA, $r_{\text{C$_2$C$_3$}} = 1.3801$ \AA, and $r_{\text{C$_3$N}} = 1.1644$ \AA.

The two dimensional PES for the HMgC$_3$N(X$^1\Sigma^+$)--He($^1$S) interaction was calculated using the explicitly correlated CCSD(T)-F12A method \citep{Knizia2009} in conjunction with the aug-cc-pVTZ basis set. The radial coordinate $R$ was sampled from 1.8 to 30.0 \AA\ using a variable step size, with a minimum spacing of 0.1 \AA. The angular coordinate $\theta$ was sampled from $0^\circ$ to $180^\circ$ in increments of $5^\circ$. This grid comprised 3626 distinct geometries, covering the short-range repulsive, the potential-well, and the long-range interaction region. The basis-set superposition error (BSSE) was corrected using the Boys--Bernardi counterpoise procedure \citep{Boys1970}:
\begin{equation}
	\label{eq.1}
	V(R,\theta ) = E_{\text{HMgC}_3\text{N--He}}(R,\theta ) - E_{\text{HMgC}_3\text{N}}(R,\theta ) - E_{\text{He}}(R,\theta ).
\end{equation}
Here, $E_{\text{HMgC}_3\text{N--He}}$, $E_{\text{HMgC}_3\text{N}}$, and $E_{\text{He}}$ represent the total energies of the HMgC$_3$N--He, HMgC$_3$N , and He, respectively.

\begin{figure}
	\centering
	\includegraphics[width=\hsize]{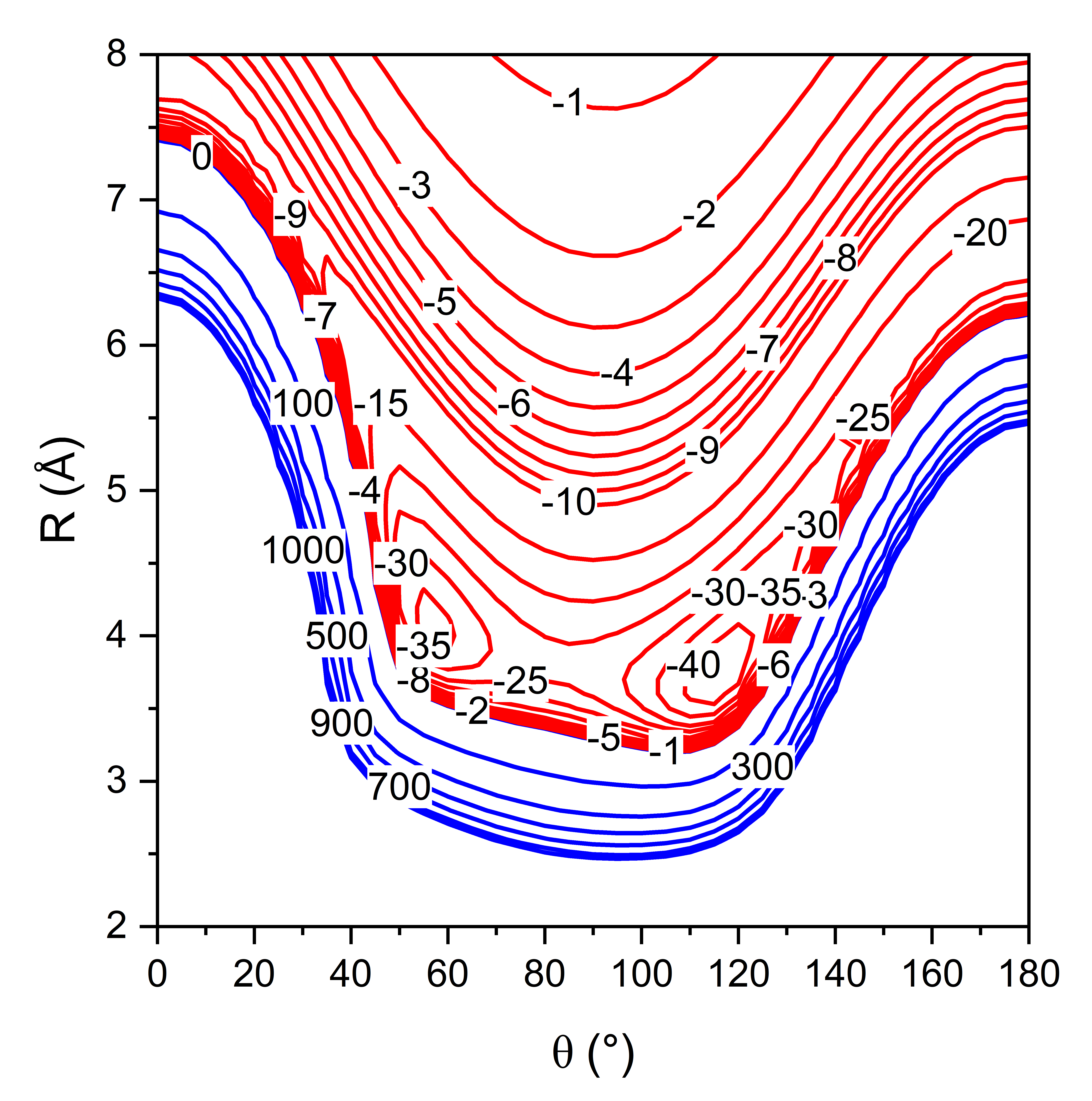}
	\caption{Contour plot  of the HMgC$_3$N--He PES as a function of Jacobi coordinates $R$ and $\theta$. The potential energy is given in cm$^{-1}$, which negative and positive values are shown in red and blue, respectively.}
	\label{fig.2}
\end{figure}

The calculated two-dimensional PES for the HMgC$_3$N--He interaction is shown in Fig.~\ref{fig.2}.
The global minimum occurs at $R = 3.74$ \AA\ and $\theta = 117.0^\circ$, with a well depth of 44.44 cm$^{-1}$. This well depth is comparable to that reported for the related MgC$_3$N--He system \citep{Khadri2025}. The PES also exhibits a second local minimum, distinct from the global minimum in both position and energy. The presence of multiple minima suggests that the anisotropy of the interaction plays an important role in shaping the PES. The calculated atomic charges of HMgC$_3$N, with the values in parentheses in Fig.~\ref{fig.1} being the results obtained in this work, indicate a non-uniform charge distribution along the molecular chain: C$_2$ carries a positive charge of \(+2.0072\), and Mg also carries a positive charge of \(+0.0321\) in the charge-partitioning scheme employed. Consequently, the electron cloud of the He atom is strongly attracted to the highly positive C$_2$ region, which well explains why the global minimum is located directly above the C$_2$ atom at \(\theta=117.0^\circ\). Moreover, the positive charge distributions at C$_2$ and Mg are consistent with the positions of the global minimum and local minimum on the PES, respectively.

\begin{figure}
	\centering
	\includegraphics[width=\hsize]{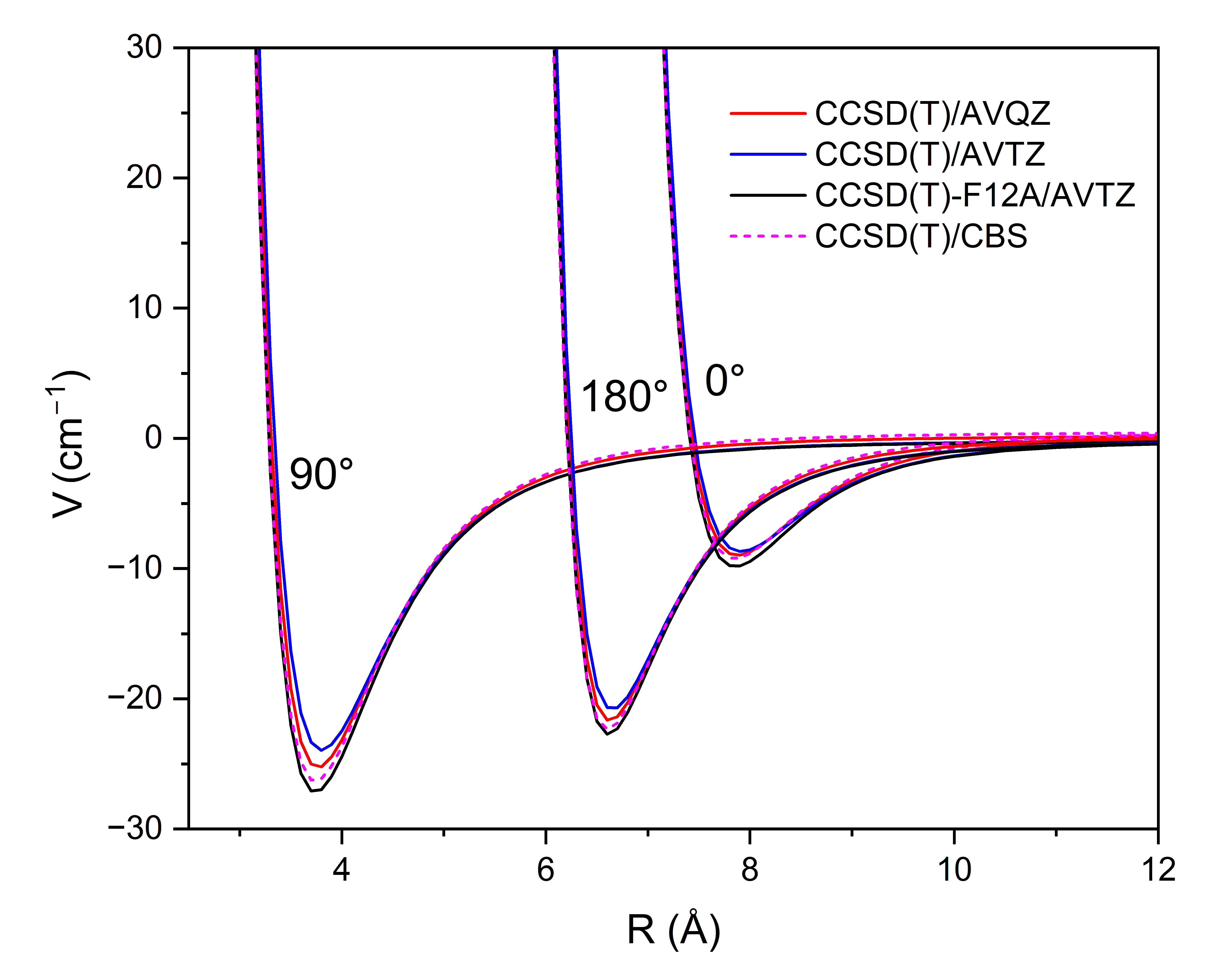}
	\caption{ Potential energy curves for HMgC3N–He at $\theta = 0^\circ$, $90^\circ$ and $180^\circ$. These curves were calculated using the CCSD-F12A/AVTZ method, as well as the CCSD(T)/aug-cc-pVXZ (X = T, Q) methods and extrapolated to the CBS limit.}
	\label{fig.3}
\end{figure}

To assess the accuracy of the CCSD(T)-F12A/AVTZ method for the HMgC$_3$N--He system, we calculated one-dimensional potential energy curves at $\theta = 0^\circ$, $90^\circ$ and $180^\circ$, respectively. The reference energies were obtained using the CCSD(T)/aug-cc-pVXZ (X = T, Q) approach. The complete basis set (CBS) limit was then estimated using a two-point extrapolation formula \citep{Peterson1994}:
\begin{equation}
	\label{eq.2}
	E_{\mathrm{CBS}} = \frac{E_X X^3 - E_Y Y^3}{X^3 - Y^3}.
\end{equation}

Here, $E_X$ and $E_Y$ denote the energies calculated with basis sets of cardinal numbers $X=4$ (AVQZ) and $Y=3$ (AVTZ), respectively.

As shown in Fig.~\ref{fig.3}, the CCSD(T)-F12A/aVTZ potential curves closely reproduce the conventional CCSD(T) results extrapolated to the CBS limit, with maximum absolute deviations below 1 cm$^{-1}$ in the attractive region. Moreover, the explicitly correlated approach reduces the computational cost by approximately two orders of magnitude compared with the CCSD(T) calculations.
All \textit{ab initio} computations in this study were carried out utilizing the {\fontsize{6}{12}\selectfont MOLPRO} program \citep{Werner2012}.

\subsection{Analytical Representation}
For  HMgC$_3$N--He system, the PES can contain very high-energy points at short intermolecular separations, often referred to as "ghost points" in the context of  fitting procedure. Including these points directly in a Legendre polynomial fit can cause the high-energy region to dominate the fitting procedure, thereby reducing the accuracy of the fit in the low-energy region that is most relevant to the scattering calculations. To mitigate this problem, we scaled the high-energy points using the procedure proposed by \cite{Faure2011}. A similar scaling procedure has previously been applied to large linear molecules, such as C$_5$S--He \citep{Khadri2020}, C$_7$N$^-$--He \citep{Gonzalez2024}. Two potential energy thresholds were introduced to control the transition from the unscaled region to the scaled high-energy region. This procedure limits the influence of the strongly repulsive region on the fit while aiming to preserve the accuracy of the PES in the low-energy region relevant to scattering calculations. The scaled potential V$_S$ computational procedure is outlined as follows:
\begin{equation}
	\label{eq.3}
	V_s =
	\begin{cases}
		V_a, & V_a \le V_{T1}, \\[4pt]
		V_{T1} + (V_{T2}-V_{T1})\,S\!\left(\dfrac{V_a-V_{T1}}{V_{T2}-V_{T1}}\right),
		& V_{T1} < V_a \le V_{T2}, \\[8pt]
		V_{T1} + \dfrac{2}{\pi}(V_{T2}-V_{T1}), & V_a > V_{T2}.
	\end{cases}
\end{equation}

Here, $V_s$ and $V_a$ denote the scaled potential energy and the original \textit{ab initio} energy, respectively. The two energy thresholds, $V_{T1}$ and $V_{T2}$, are set to 2000 and 20000 cm$^{-1}$ in the present work. The switching function is defined as $S(x) = \frac{2}{\pi} \sin \left[ \frac{\pi}{2} \sin \left( \frac{\pi}{2} x \right) \right]$ . This function provides a continuous transition in the potential energy between the intermediate and high energy regions.

The  PES obtained from the \textit{ab initio} calculations was represented using a Legendre polynomial expansion \citet{Werner1988}:
\begin{equation}
	\label{eq.4}
	\ V(R,\theta )=\sum\limits_{\lambda =0}^{{{\lambda }_{\max }}}{{{V}_{\lambda }}}(R){{P}_{\lambda }}(\cos \theta ) \,
\end{equation}
where $P_\lambda(\cos\theta)$ is the Legendre polynomial of order $\lambda$ and \(V_\lambda(R)\) is the corresponding radial expansion coefficient. The expansion was truncated at \(\lambda_{\max}=58\), yielding a fitting error of 0.5 per cent according to the error metric defined for the PES fit. 

\begin{figure}
	\centering
	\includegraphics[width=\hsize]{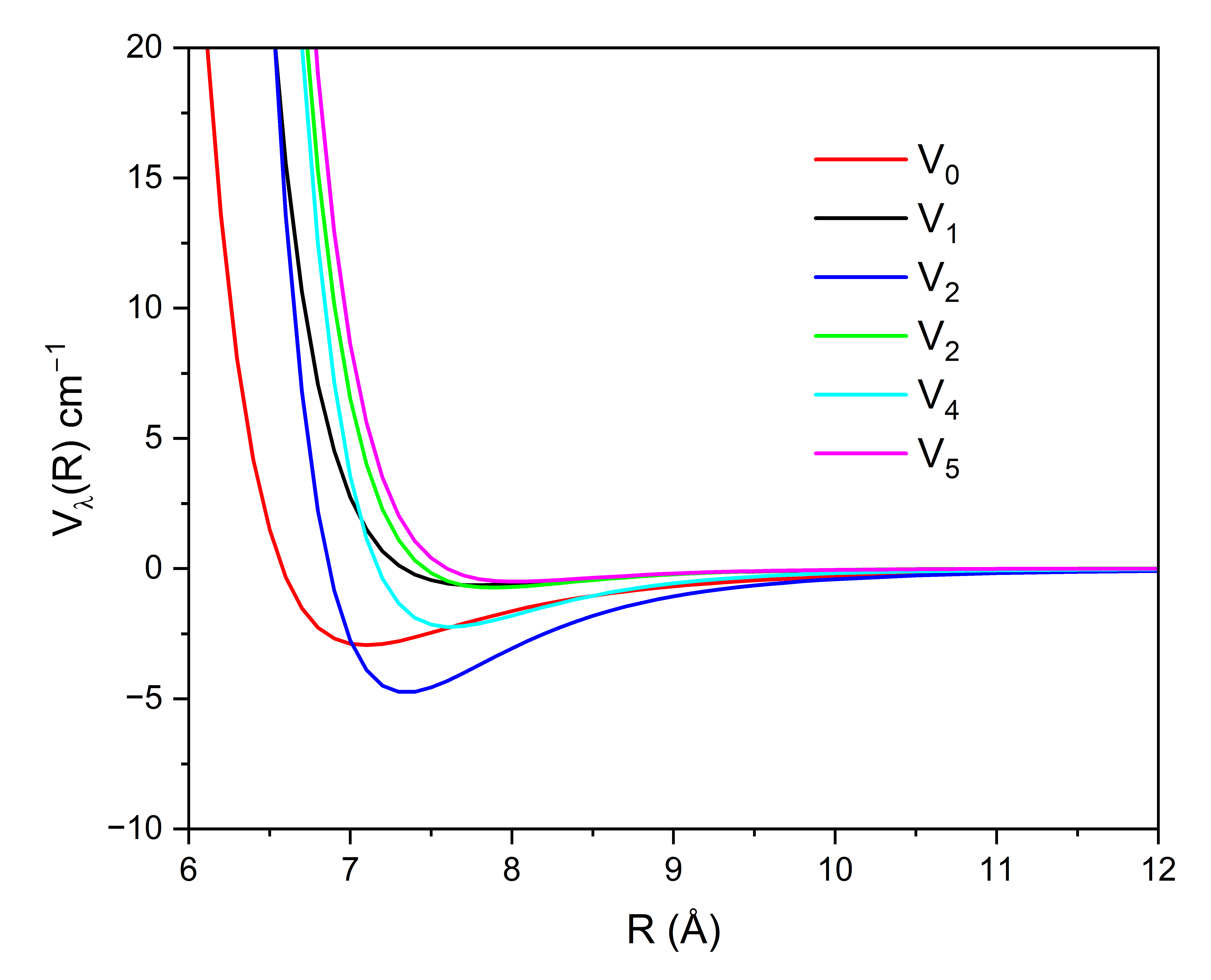}
	\caption{Radial coefficients $V_\lambda$ (in cm$^{-1}$) for $\lambda=0$--4 versus the distance $R$ for the HMgC$_3$N--He system.}
	\label{fig.4}
\end{figure}

As shown in Fig.~\ref{fig.4}, among all the anisotropic expansion terms ($\lambda > 0$), the $V_2$ term (corresponding to $\lambda = 2$) exhibits the largest amplitude.  The \(\lambda=2\) term contributes to rotational coupling between channels with \(j\) and \(j'\) satisfying the angular-momentum selection rules. Its relatively large amplitude therefore suggests that it may play an important role in the rotationally inelastic scattering dynamics.

\section{Scattering calculations and collisional properties}
\label{Scat cal}
\subsection{Cross sections}
\begin{figure*}
	\centering
	\includegraphics[width=\hsize]{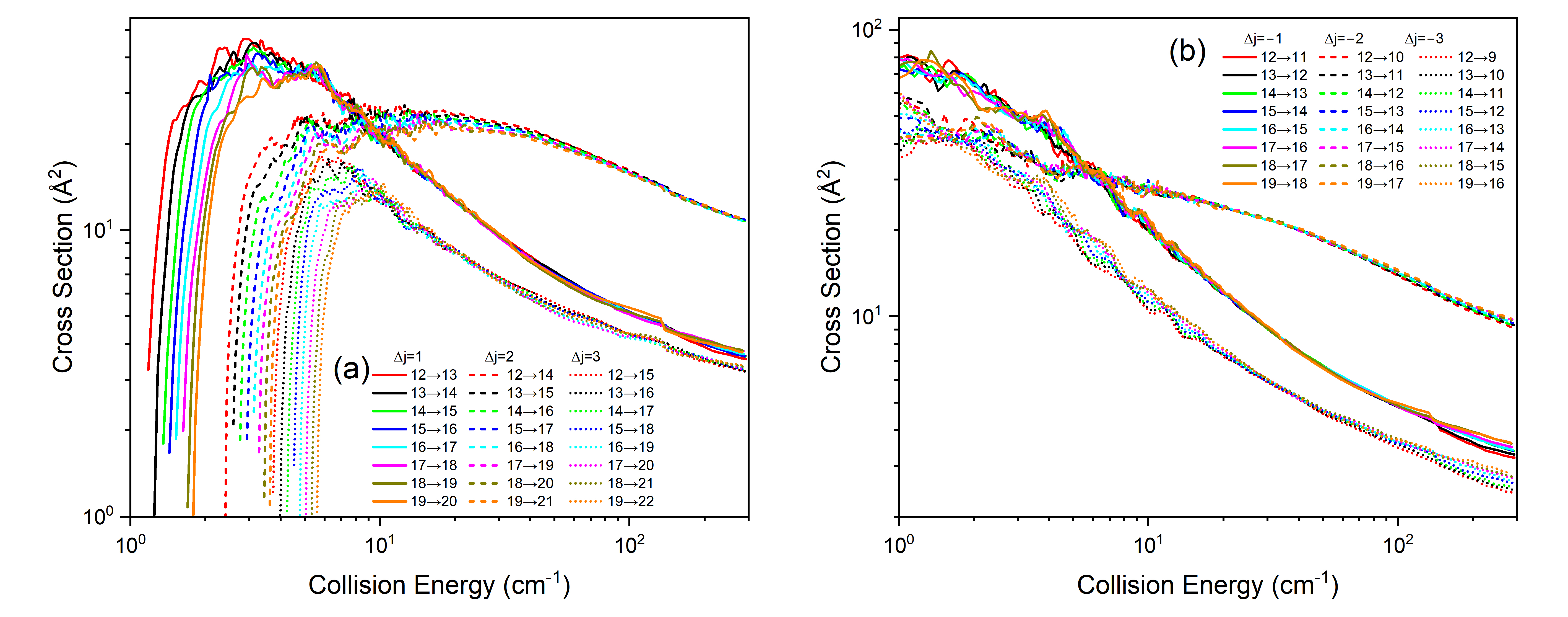}
	\caption{Cross sections (in units of \AA$^2$ ) as functions of collision energies for rotational (a) excitation and (b) de-excitation of HMgC$_3$N--He.}
	\label{fig.5}
\end{figure*}

Following the construction of the PES, we employed the CC method to calculate the state-to-state integral cross sections. For HMgC$_3$N, the rotational constant $B_0 = 0.044021$ cm$^{-1}$ and the centrifugal distortion constant $D_0 = 3.3 \times 10^{-9}$ cm$^{-1}$ were taken from the work of \citet{Cabezas2023}. The cross sections for transitions from an initial rotational level $j$ to a final level $j'$ were calculated following the formalism of \citet{Green1975} :

\begin{equation}
	\label{eq.5}
	\begin{split}
		\sigma_{j \to j'}(E_k) ={}& \frac{\pi}{k_j^2 (2j+1)} \sum_{J=0} \sum_{l=|J-j|}^{J+j} \sum_{l'=|J-j'|}^{J+j'} (2J+1) \\
		&\times \left| \delta_{jj'} \delta_{ll'} - S_{jj'll'}^{J}(E) \right|^2 .
	\end{split}
\end{equation}

Here, $k_j = \sqrt{2\mu(E - E_j)} / \hbar$ is the wave number  in the incident channel, where $\mu$
is the reduced mass, $E$ is the total energy of the system, and $E_j$ corresponds to the rotational energy of the initial state. The collision energy is therefore given by $E_k$=$E$--$E_j$. The following equation gives the rotational energy level of HMgC$_3$N.
\begin{equation}
	\label{eq.6}
	\ E_j = B_0\,j(j+1) - D_0\,[j(j+1)]^2 \, .
\end{equation}
For the HMgC$_3$N–He collision system, $\mu = 3.799804$ amu. The orbital angular momenta in the initial and final channels are denoted by $l$ and $l’$, respectively, while $J$ denotes the total angular momentum. 

The total energy grid was discretised using variable step sizes to adequately resolve resonance features: 0.05 cm$^{-1}$ over the range of  0.05--30 cm$^{-1}$, 0.1 cm$^{-1}$ over 30--50 cm$^{-1}$, 0.5 cm$^{-1}$ over 50--100 cm$^{-1}$, 1.0 cm$^{-1}$ over 100--150 cm$^{-1}$ and 5.0 cm$^{-1}$ over 150--300 cm$^{-1}$. To ensure convergence of the cross sections for transitions involving rotational levels up to
$j \le 20$, the rotational basis set truncation parameter, $j_{\text{MAX}}$, was systematically optimised different energy ranges. Specifically, $j_{\text{MAX}}$ was set to 30 for $E \le 50$ cm$^{-1}$, 32 for $50 < E \le 150$ cm$^{-1}$, and 36 for $150 < E \le 300$ cm$^{-1}$. The convergence tests confirmed that the optimised $j_{\text{MAX}}$ values yielded cross sections with numerical uncertainties of less than 5\% for the transitions considered.
Furthermore, the radial propagation was integrated from $R_{\text{MIN}} = 1.8$ \AA\ to $R_{\text{MAX}} = 20.0$ \AA. For \(E \le 50~\mathrm{cm}^{-1}\), \(50 < E \le 100~\mathrm{cm}^{-1}\), and \(E > 100~\mathrm{cm}^{-1}\), the number of integration steps per de Broglie wavelength was fixed at 30, 20, and 10, respectively. At each energy point, the total angular momentum $J_{\text{tot}}$,  controlled by the  \texttt{JTOTU},  was increased iteratively until convergence was achieved, with an upper limit of 99999. The convergence thresholds for the diagonal and off-diagonal matrix elements, DTOL and OTOL, were set to 0.1 \AA$^2$ and 0.005 \AA$^2$, respectively. Finally, the coupled equations were solved utilizing the improved log-derivative method \citep{Manolopoulos1986} in the short-range region and the Airy propagator \citep{Alexander1987} in the long-range region. All quantum scattering computations were carried out using the {\fontsize{6}{12}\selectfont MOLSCAT} software package \citep{Hutson2019}.

\begin{figure*}
	\centering
	\includegraphics[width=\hsize]{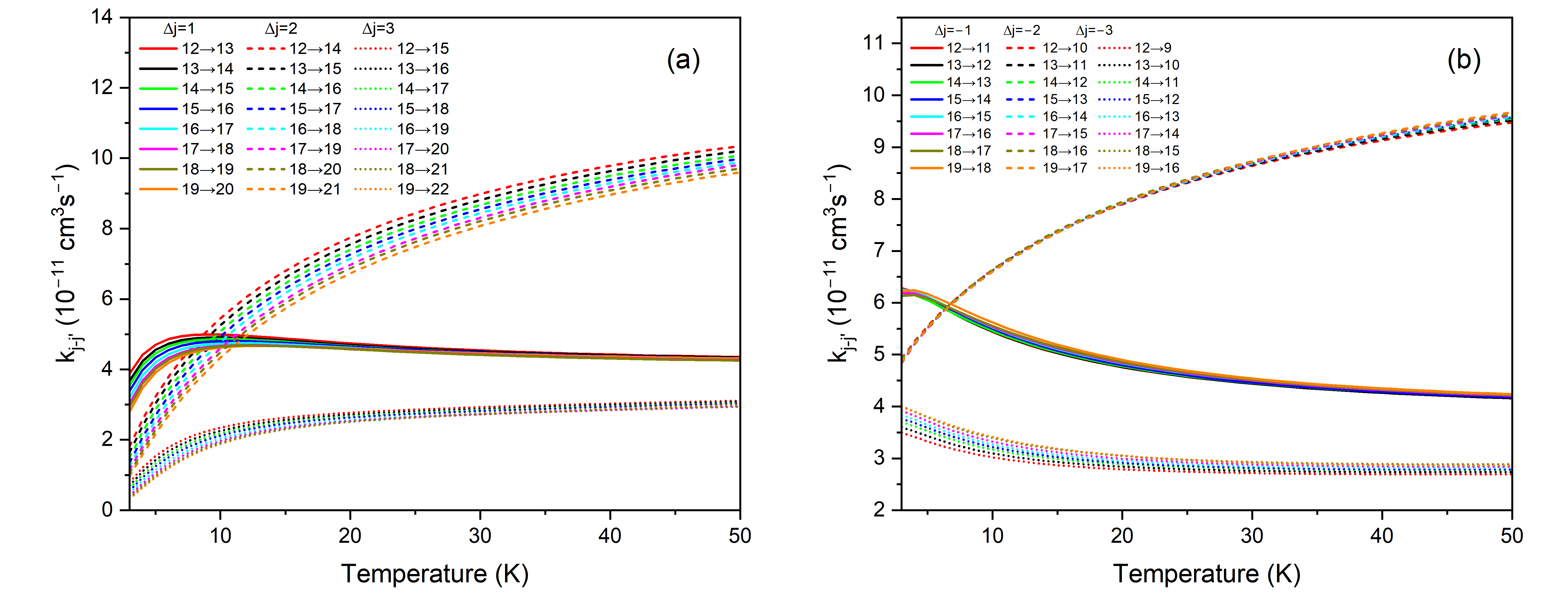}
	\caption{Rate coefficients as functions of temperature for rotational (a) excitation and (b) de-excitation of HMgC$_3$N in collisions with He.}
	\label{fig.6}
\end{figure*}

Fig.~\ref{fig.5} shows the integral cross sections for the HMgC$_3$N--He collisional system. The panels present transitions originating from initial rotational levels \(j = 12\) to \(19\), with solid, dashed, and dotted lines denoting transition channels of \(\Delta j = \pm 1, \pm 2,\) and \(\pm 3\), respectively. 
In panel (a), the excitation cross sections exhibit a dense resonance structure at low collision energies (below 10 cm$^{-1}$), which is characteristic of low energy molecular collisions with He. These resonances can be attributed to quasi-bound states supported by the PES, including shape and Feshbach resonances. As the collision energy increases, these resonance features gradually diminish, and the cross sections become smoother and generally decrease with increasing collision energy. Notably, for a given initial level, the transition of $\Delta j = +2$ becomes increasingly dominant at higher energies, followed by $\Delta j = +1$ and $\Delta j = +3$, indicating a pronounced propensity for even $\Delta j$ transitions. Furthermore, the cross sections for all excitation channels become relatively smooth in the high-energy region.
Panel (b) illustrates the de-excitation process, where pronounced low-energy resonances are similarly observed. At higher energies, a distinct even--odd propensity becomes evident: over most of the energy range, the cross sections for $\Delta j = -2$ transitions (e.g., $12 \to 10$, $18 \to 16$) are substantially larger than those for $\Delta j = -1$ transitions (e.g., $12 \to 11$, $18 \to 17$). This propensity is consistent with the dominance of the $\lambda = 2$ anisotropic term in the interaction potential and agrees with the transition propensity discussed above. Additionally, for a given $\Delta j$, the de-excitation cross sections generally increase with the initial rotational level $j$.  However, for transitions between highly excited states (e.g., $18 \to 17$ and $19 \to 18$), the variations in magnitude become negligible. These systematic trends highlight the important role of the PES anisotropy in governing rotational energy transfer.
It should be noted that, for the high-energy scattering cross sections calculations, \texttt{JSTEP} in was increased from 1 to 2 or 3 to accelerate the computations; the resulting oscillations in the high-energy tails of the cross section curves (above \(150~\mathrm{cm}^{-1}\)) are numerical artifacts and should not be interpreted as physical resonance features.

\subsection{Rate coefficients}

The rotational (de-)excitation rate coefficients were obtained by thermal averaging of the cross sections:
\begin{equation}
	\label{eq.7}
	{{k}_{j\to {j}'}}(T)={{\left( \frac{8}{\pi \mu k_{\text{B}}^{3}{{T}^{3}}} \right)}^{1/2}}\int_{0}^{\infty }{{{E}_{k}}}{{\sigma }_{j\to {j}'}}({{E}_{k}}){{e}^{-{{E}_{k}}/{{k}_{\text{B}}}T}}d{{E}_{k}}.
\end{equation}

Fig.~\ref{fig.6}(a) and Fig.~\ref{fig.6}(b) display the temperature dependence of the excitation and de-excitation rate coefficients in the HMgC$_3$N--He system over the 3--50 K range, respectively. In panel (a), at low temperatures ($T < 10$ K), the $\Delta j = +1$ channels dominate. As the temperature increase, the rate coefficients for the  $\Delta j = +2$ transitions exhibit a pronounced monotonic increases, with the rate coefficient for the $12 \to 14$ transition
reaching approximately $1.0 \times 10^{-10}$ cm$^3$ s$^{-1}$ at 50 ~K. Additionally, for a fixed $\Delta j$, the rate coefficients systematically decrease as the initial rotational level $j$ increases.
Similar to the excitation rates, the de-excitation rate coefficients exhibit a clear dependence on the initial rotational quantum number $j$. For a given $\Delta j$, the temperature dependence varies with the initial rotational level. At temperatures above 10 ~K, a distinct propensity becomes evident: for each initial rotational level $j$, the rate coefficients for $\Delta j = -2$ transitions are consistently larger than those for the corresponding  $\Delta j = -1$ transitions.
This preference for even-$\Delta j$ transitions during de-excitation is consistent with the cross section patterns and further supports the important role of the even-order anisotropic components of the PES. Notably, for both excitation and de-excitation processes, the rates for \(|\Delta j| = 3\) remain lower than those for \(|\Delta j| = 1\), 2.
Overall, the rate coefficients vary smoothly with temperature because the dense resonance features present in the cross sections are effectively averaged out during thermal averaging. This smooth temperature dependence reflects the thermal averaging of the underlying collision dynamics and provides essential collisional data for non-LTE modelling of HMgC$_3$N in astrophysical environments.

\section{Astrophysical applications}
\begin{figure*}
	\centering
	\includegraphics[width=\hsize]{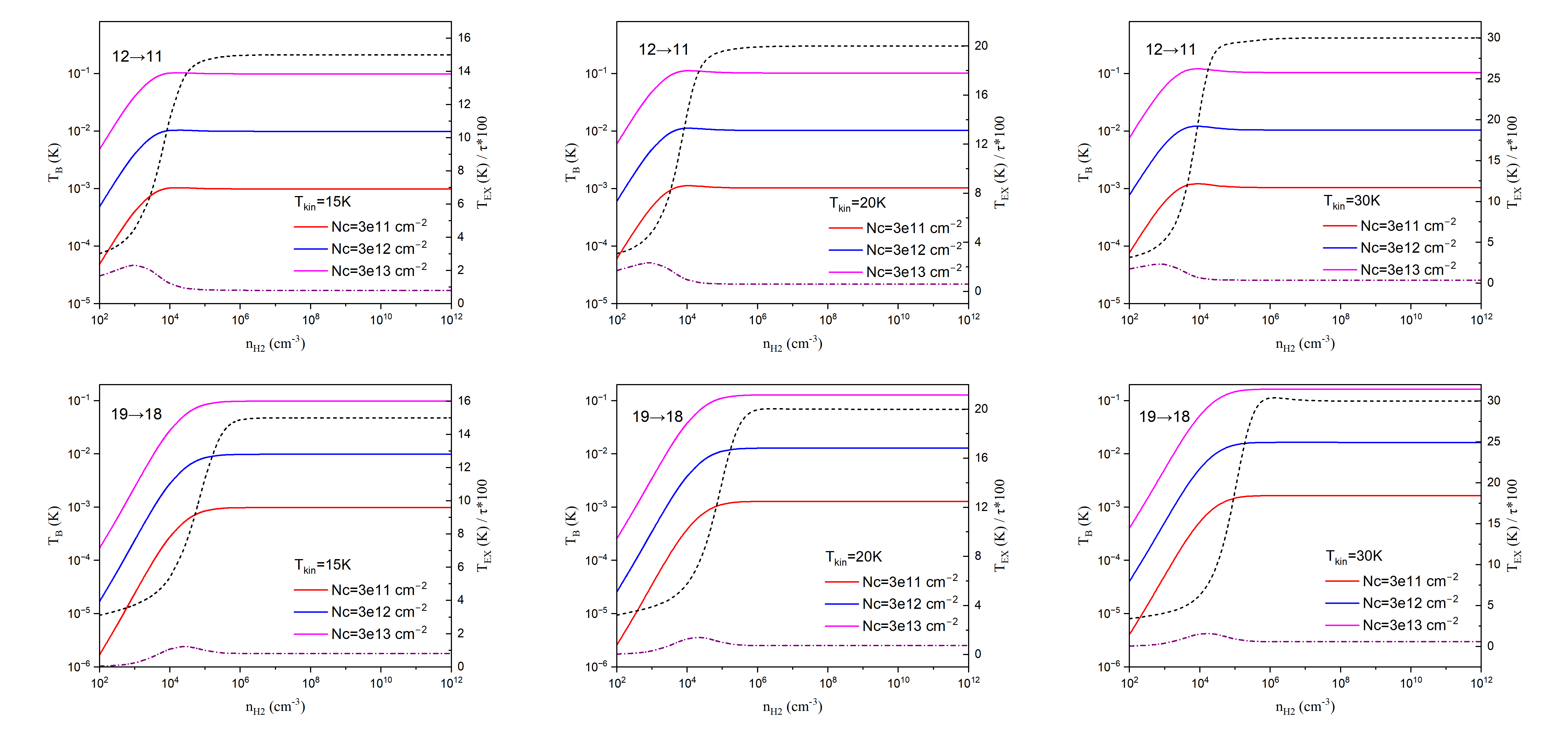}
	\caption{For three different HMgC$_3$N column densities, the excitation temperatures (black dashed lines) and brightness temperatures (coloured curves) corresponding to the $12\rightarrow11$ and $19\rightarrow18$ transitions of HMgC$_3$N are shown as a function of H$_2$ density. The assumed kinetic temperatures are 15 K, 20 K, and 30 K, respectively. The optical depth is scaled by a factor of 100.}
	\label{fig.7}
\end{figure*}
\label{sec:astro}
\begin{table}
	\caption{Critical densities $n_{\mathrm{c}}$ ($\mathrm{cm}^{-3}$) of
		HMgC$_3$N at 15, 20, and 30\,K for the observed transitions.}
	\label{tab:1}
	\centering
	\small
	\setlength{\tabcolsep}{6pt}
	\begin{tabular}{lccc}
		\hline\hline
		$J\rightarrow J'$ & 15\,K & 20\,K & 30\,K \\
		\hline
		$12\rightarrow11$ & $1.3925\times10^{4}$ & $1.3801\times10^{4}$ & $1.3618\times10^{4}$ \\
		$13\rightarrow12$ & $1.7245\times10^{4}$ & $1.7073\times10^{4}$ & $1.6815\times10^{4}$ \\
		$14\rightarrow13$ & $2.1031\times10^{4}$ & $2.0800\times10^{4}$ & $2.0434\times10^{4}$ \\
		$15\rightarrow14$ & $2.5341\times10^{4}$ & $2.5062\times10^{4}$ & $2.4601\times10^{4}$ \\
		$16\rightarrow15$ & $3.0137\times10^{4}$ & $2.9778\times10^{4}$ & $2.9163\times10^{4}$ \\
		$17\rightarrow16$ & $3.5559\times10^{4}$ & $3.5144\times10^{4}$ & $3.4415\times10^{4}$ \\
		$18\rightarrow17$ & $4.1463\times10^{4}$ & $4.0942\times10^{4}$ & $4.0010\times10^{4}$ \\
		$19\rightarrow18$ & $4.8104\times10^{4}$ & $4.7525\times10^{4}$ & $4.6441\times10^{4}$ \\
		\hline
	\end{tabular}
\end{table}

Since H$_2$ is the dominant collision partner in interstellar molecular clouds, reliable astrophysical modelling requires accurate collisional rate coefficients for the HMgC$_3$N--H$_2$ system. In the absence of dedicated HMgC$_3$N--H$_2$ collisional data, we estimated these coefficients by applying a reduced-mass scaling approximation to the previously calculated HMgC$_3$N--He rate coefficients, using a scaling factor of 1.39:

\begin{equation}
	k_{\mathrm{HMgC_3N-H_2}}(T) =
	\sqrt{\frac{\mu_{\mathrm{HMgC_3N-He}}}{\mu_{\mathrm{HMgC_3N-H_2}}}} \times
	k_{\mathrm{HMgC_3N-He}}(T) .
\end{equation}
It should be noted that this scaling method is only a practical approximation and does not imply that the two collision systems are equivalent. Previous studies have shown that rate coefficients for collisions with H$_2$ can differ from those for collisions with He by factors of several, depending on the molecular species and transitions involved \citep{Lique2006, Denis2013}. In the subsequent non-LTE radiative transfer calculations, we therefore adopted the reduced-mass-scaled coefficients as approximate H$_2$ collisional rate coefficients.

In astrophysical environments, the establishment of a LTE distribution of molecular level populations is governed by the competition between collisional and radiative processes. The critical density, $n_{\mathrm{c}}$, is a characteristic parameter for assessing the relative importance of collisional de-excitation and spontaneous radiative decay. Following the formalism of \citet{lique2018}, we evaluated this quantity as

\begin{equation}
	n_{\mathrm{crit}}(j) =
	\frac{A_{j \to j-1}}{\displaystyle\sum_{j'<j} k_{j \to j'}}.
\end{equation}

Here, the critical density is defined as the ratio of the Einstein $A$ coefficient for the $j\to j-1$ transition to the sum of the collisional de-excitation rate coefficients from the initial level $j$ to all lower rotational levels \citep{Lique2009, Lique2014}. Motivated by the detection of HMgC$_3$N in IRC+10216 \citep{Cabezas2023}, we focus on the rotational transitions with upper-state quantum numbers $j_{\text{up}}=12$--19 covered by these observations. We calculated the Einstein $A$ coefficients for the $\Delta j=-1$ transitions within this range; these values are provided as online supplementary data (see the Data Availability section). The resulting critical densities at kinetic temperatures ($T_{\mathrm{kin}}$) of 15, 20, and 30 K are summarised in Table~\ref{tab:1}.

The radiative transfer calculations were performed using the RADEX code \citep{van2007}. We specifically examined the lowest- and highest-$j$ transitions in the sample observed by \citet{Cabezas2023}, namely $12\to11$ and $19\to18$, respectively. The corresponding results are presented in Fig.~\ref{fig.7}. 
In the low-density regime, the excitation temperatures of all transitions initially approach the cosmic microwave background temperature and begin to increase significantly at H$_2$ densities between approximately $10^3$ and $10^4$ cm$^{-3}$. This density range is comparable to the critical densities listed in Table~\ref{tab:1}, which are of the order of $10^4$ cm$^{-3}$. As the H$_2$ density increases further, the excitation temperatures gradually level off and approach the corresponding $T_{\text{kin}}$, indicating thermalised excitation. As shown in the Fig.~\ref{fig.7}, the excitation temperatures of the transitions considered here approach their LTE limits at densities of $n(\mathrm{H}_2)\geq10^5$ cm$^{-3}$.

Meanwhile, the brightness temperature $T_{\mathrm{B}}$ increases systematically with the adopted column density. Notably, the $T_{\mathrm{B}}$ curves for the lower-energy $12\to11$ transition reach a stable plateau at lower densities than those for the higher-energy $19\to18$ transition. Specifically, at all three kinetic temperatures, the brightness temperature of the $12\to11$ line stabilises at an H$_2$ density of approximately $10^4$ cm$^{-3}$, whereas the $19\to18$ transition requires a density of approximately $10^5$ cm$^{-3}$ to reach a comparable plateau. This behaviour indicates that higher-$j$ transitions generally require higher H$_2$ densities for their excitation to become thermalised. A similar trend has been reported for the SiC$_4$--He system \citep{Chefai2024}, where the density required to thermalise the $28\to27$ transition is notably higher than that for lower-$j$ transitions. Furthermore, the lower panels show that \(\tau \leq 0.03\) for all emission lines, confirming that the modelled emission lines are optically thin.

Although higher densities, exceeding $10^6$ cm$^{-3}$, may be required to bring the entire rotational level distribution closer to complete thermalization, the physical conditions of the emitting region can be constrained by fitting the observed high-$j$ spectral lines. To quantitatively assess the effects of non-LTE excitation and examine the column density derived from the observations, we performed a $\chi^2$ minimization procedure \citep{Lampton1976}.
\begin{equation}
	\chi^2 = \sum_{i=1}^{N}
	\left(
	\frac{I_{\mathrm{calc},i}-I_{\mathrm{obs},i}}
	{\sigma_{I,i}}
	\right)^2
\end{equation}

Here, $I_{\mathrm{calc},i}$ and $I_{\mathrm{obs},i}$ denote the synthetic and observed integrated intensities of the $i$-th transition, respectively, while $\sigma_{I,i}$ represents the corresponding observational uncertainty. The reduced chi-square is defined as $\chi^2_{\mathrm{min}}=\chi^2/(N-p)$, where $N$ is the number of fitted data points and $p$ is the number of free fitting parameters.

\begin{figure}
	\centering
	\includegraphics[width=\hsize]{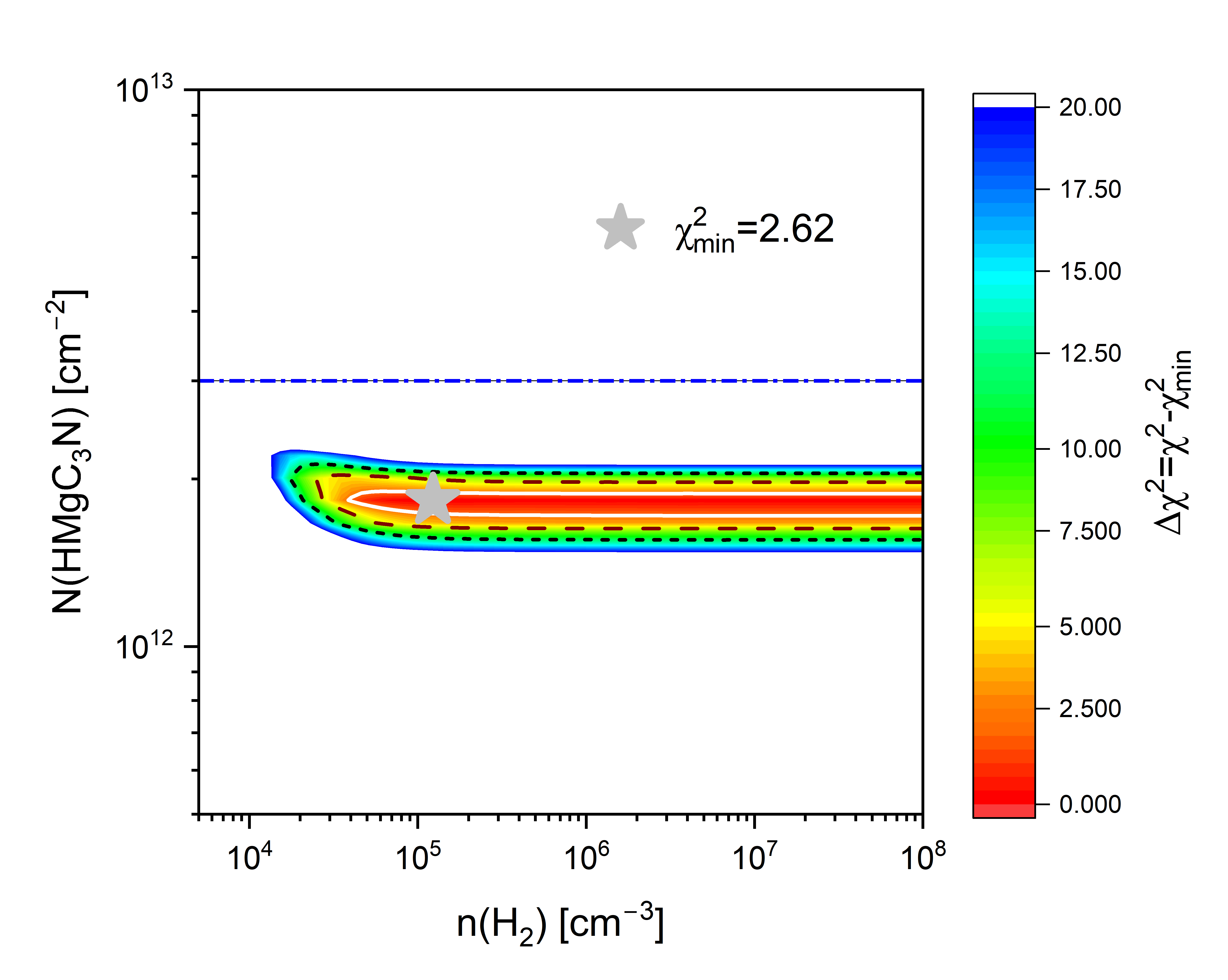}
	\caption{Two-dimensional $\chi^2$ minimization grid as a function of the H$_2$ volume density, $n(\mathrm{H}_2)$, and the HMgC$_3$N column density, $N(\mathrm{HMgC}_3\mathrm{N})$. The global minimum ($\chi^2_{\min} = 2.62$) is marked by the red star. The solid white, dashed wine, and short-dashed black contours denote the $1\sigma$, $2\sigma$, and $3\sigma$ confidence levels ($\Delta\chi^2 = 2.30, 6.18, 11.83$), respectively.}
	\label{fig.8}
\end{figure}

Fig.~\ref{fig.8} shows the two-dimensional $\chi^2$ minimization grid obtained from the non-LTE radiative transfer modelling, which was used to assess the constraints provided by the fitting procedure. The horizontal and vertical axes represent the H$_2$ volume density, $n(\mathrm{H}_2)$, and the HMgC$_3$N column density, $N(\mathrm{HMgC}_3\mathrm{N})$, respectively. The colour scale, ranging from 0 to 20, represents $\Delta\chi^2$. The star symbol marks the global minimum, corresponding to a reduced chi-square of \(\chi^2_{\text{min}}=2.62\). The contour distribution exhibits features characteristic of sub-thermal excitation: the confidence region is highly elongated along the density axis but relatively narrow along the column-density axis. This indicates a pronounced degeneracy between the $n(\mathrm{H}_2)$ and $N(\mathrm{HMgC}_3\mathrm{N})$, with the modelled line intensities being less sensitive to variations in density than to variations in column density. Nevertheless, the column density remains relatively well constrained. The short blue dashed line indicates the LTE-derived column density reported in the literature ($3\times10^{12}$ cm$^{-2}$), which lies slightly above the best-fit value but within the confidence region of the non-LTE model.

The best-fit H$_2$ density derived from the six observed transitions is $1.233\times10^5$ cm$^{-3}$, approximately one order of magnitude higher than the typical ambient density in the envelope of IRC+10216 ($\sim10^4$ cm$^{-3}$) \citep{Cabezas2023}. As shown in Fig.~\ref{fig.7}, at this density, the excitation temperature of the higher-energy $19\to18$ transition has not yet fully reached its plateau, whereas that of the $12\to11$ transition is already approaching the LTE limit. Although the $1\sigma$ and $2\sigma$ contours reveal a pronounced degeneracy between $n(\mathrm{H}_2)$ and $N(\mathrm{HMgC}_3\mathrm{N})$, the adopted modelling framework and fitting procedure still provide meaningful constraints on the column density.

\begin{figure}
	\centering
	\includegraphics[width=\hsize]{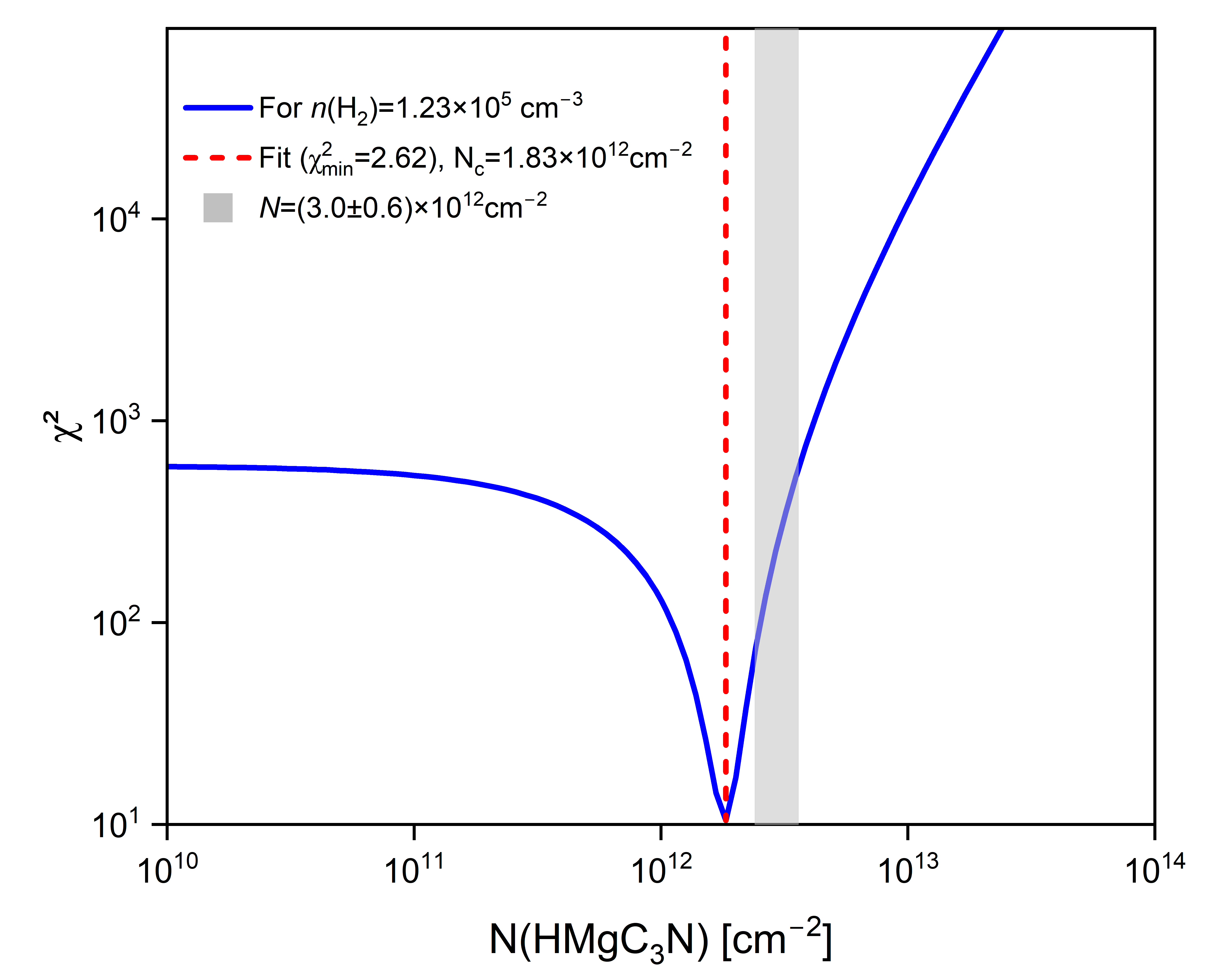}
	\caption{One-dimensional $\chi^2$ profile evaluated at $n(\mathrm{H}_2) = 1.233 \times 10^5~\mathrm{cm}^{-3}$ as a function of $N(\mathrm{HMgC}_3\mathrm{N})$. The vertical red dashed line indicates our optimal fit ($N = 1.831 \times 10^{12}~\mathrm{cm}^{-2}$). The light gray shaded area indicates the values from the study by \citet{Cabezas2023}.}
	\label{fig.9}
\end{figure}

Fig.~\ref{fig.9} shows the one-dimensional $\chi^2$ profile as a function of $N(\mathrm{HMgC}_3\mathrm{N})$ at the best-fit H$_2$ density of $n(\mathrm{H}_2)=1.23\times10^5$ cm$^{-3}$. This analysis was performed to quantify the constraint on the column density and assess potential biases associated with the LTE assumption. The solid blue curve shows the variation of $\chi^2$ with column density, reaching its minimum (indicated by the short red dashed line) at $N(\mathrm{HMgC}3\mathrm{N})=1.83\times10^{12}$ cm$^{-2}$, corresponding to $\chi^2{\mathrm{min}}=2.62$. The gray shaded band indicates the reference column density and its uncertainty range, $N=(3.0\pm0.6)\times10^{12}$ cm$^{-2}$, derived from the conventional LTE rotational diagram. The best-fit non-LTE column density is approximately a factor of 1.6 lower than the LTE estimate. This difference is considerably smaller than those reported for some other molecular systems exhibiting substantial departures from LTE, such as the HC$_4$NC--He system \citep{Terzi2026}, for which the difference exceeds a factor of 6. These results indicate that the observed HMgC$_3$N transitions exhibit some degree of sub-thermal excitation. Nevertheless, the LTE-derived column density remains consistent with the non-LTE result in order of magnitude, although the non-LTE best-fit value falls below the $1\sigma$ uncertainty interval of the LTE estimate.

\section{Discussion and conclusions}
\label{sec:conclusions}
This paper investigated the rotationally inelastic collisions between HMgC$_3$N with He. A two-dimensional PES was constructed by CCSD(T)-F12A/AVTZ method, with a global minimum of 44.44 cm$^{-1}$ at $R = 3.74$ \AA\ and $\theta = 117.0^\circ$. CC method was performed to obtain state-to-state inelastic integral cross sections and rate coefficients for rotational transitions up to $j=20$ over the temperature range of 3--50 K. The calculated rate coefficients exhibit a propensity for even-$\Delta j$ transitions, mainly associated with the dominant $\lambda=2$ component of the PES.

Radiative transfer calculations were also performed to model the observed HMgC$_3$N emission lines toward IRC+10216. In the absence of HMgC$_3$N--H$_2$ data, the corresponding rate coefficients were estimated by scaling the HMgC$_3$N--He rates according to the reduced masses. A $\chi^2$ minimization analysis of six observed transitions yielded best-fit values of $n(\mathrm{H}_2)=1.233\times10^5$ cm$^{-3}$ and $N(\mathrm{HMgC}_3\mathrm{N})=1.831\times10^{12}$ cm$^{-2}$. The non-LTE column density is of the same order of magnitude as, but lower than, the value derived from the LTE analysis. The calculated collisional rate coefficients will be made publicly available through the LAMDA database to facilitate future non-LTE studies of interstellar HMgC$_3$N.

\section*{Data availability}
The collisional rate coefficients and the associated molecular data required for the RADEX calculations will be deposited in the Zenodo repository and submitted to the LAMDA database upon acceptance for publication. 

\section*{Acknowledgements}

This research did not receive any specific grant from funding agencies in the public, commercial, or not-for-profit sectors.



\bibliographystyle{mnras}
\bibliography{example} 










\bsp	
\label{lastpage}
\end{document}